\documentclass[iop,numberedappendix,twocolappendix]{emulateapj}

\usepackage{natbib}
\usepackage{amsmath}
\usepackage{graphicx}
\usepackage{dcolumn}
\usepackage{bm}
\renewcommand{\phi}{\varphi}
\renewcommand{\theta}{\vartheta}

\newcommand{\kten}{\ensuremath{\hat{\kappa}}}

\shorttitle{A Generalized Diffusion Tensor for Fully Anisotropic Diffusion}
\shortauthors{Effenberger et al.}

\begin{document}

\title{A Generalized Diffusion Tensor for Fully Anisotropic Diffusion of Energetic Particles in the Heliospheric Magnetic Field}

\author{F. Effenberger\altaffilmark{1}, H.
  Fichtner\altaffilmark{1}, K. Scherer\altaffilmark{1}, S.
  Barra\altaffilmark{1}, J. Kleimann\altaffilmark{1}, and R. D. Strauss\altaffilmark{2}} 
\altaffiltext{1}{Institut f\"ur
  Theoretische Physik IV, Ruhr-Universit\"at Bochum, 44780 Bochum,
  Germany, Email: fe@tp4.rub.de}
\altaffiltext{2}{Centre for Space Research, North-West University, 2520 Potchefstroom, South Africa}

\begin{abstract}
  The spatial diffusion of cosmic rays in turbulent magnetic fields
  can, in the most general case, be fully anisotropic, i.e.\ one has
  to distinguish three diffusion axes in a local, field-aligned
  frame. We reexamine the transformation for the diffusion tensor from
  this local to a global frame, in which the Parker transport equation
  for energetic particles is usually formulated and
  solved. Particularly, we generalize the transformation formulas to
  allow for an explicit choice of two principal local perpendicular
  diffusion axes. This generalization includes the 'traditional'
  diffusion tensor in the special case of isotropic perpendicular
  diffusion. For the local frame, we motivate the choice of the
  Frenet-Serret trihedron which is related to the intrinsic magnetic
  field geometry. We directly compare the old and the new tensor
  elements for two heliospheric magnetic field configurations, namely
  the hybrid Fisk and the Parker field. Subsequently, we examine the
  significance of the different formulations for the diffusion tensor
  in a standard 3D model for the modulation of galactic protons. For
  this we utilize a numerical code to evaluate a system of stochastic
  differential equations equivalent to the Parker transport equation
  and present the resulting modulated spectra. The computed
  differential fluxes based on the new tensor formulation deviate from
  those obtained with the 'traditional' one (only valid for isotropic
  perpendicular diffusion) by up to 60\% for energies below a few
  hundred MeV depending on heliocentric distance.
\end{abstract}


\section{Introduction and motivation}
The most important transport process for energetic charged particles
in the heliosphere is their spatial diffusion as a consequence of
their interaction with the turbulent heliospheric magnetic field
(HMF). While rarely used in models of galactic transport, the concept
of {\it anisotropic diffusion} is well established \citep[see,
e.g.,][]{Burger-etal-2000,Schlickeiser-2002,Shalchi-2009} for models
of heliospheric cosmic ray (CR) modulation. In most studies the
anisotropy refers to a difference in the diffusion coefficient
parallel ($\kappa_{\parallel}$) and perpendicular ($\kappa_{\perp}$)
to the magnetic field, and the perpendicular diffusion is treated as
being isotropic.

Although the notion of fully anisotropic diffusion is not 
new -- see an early study by \citet{Jokipii-1973} considering for the first
time {\it anisotropic perpendicular diffusion}, i.e.\ $\kappa_{\perp 1} \neq 
\kappa_{\perp 2}$ -- it was not before the measurements made with the Ulysses
spacecraft that this concept had to be used to explain the high-latitude 
observations of cosmic rays, see, e.g., \citet{Jokipii-etal-1995}, 
\citet{Potgieter-etal-1997}, and \citet{Ferreira-etal-2001}. These studies
remained largely phenomenological and did not attempt a rigorous
investigation of anisotropic perpendicular diffusion.  

More recently, in the context of studies of the transport of solar
energetic particles in the heliospheric Parker field,
\citet{Tautz-etal-2011} and \citet{Kelly-etal-2012} have determined
the elements of the diffusion tensor from test particle simulations in
a local, field-aligned frame. While the former authors find no
conclusive result, the latter authors clearly demonstrated that the
scattering in the inhomogeneous Parker field can indeed induce
anisotropic perpendicular diffusion.

Given this phenomenological and simulation-based evidence it is
important to determine the principal directions of perpendicular
diffusion in the field-aligned local frame, because the transformation
of the diffusion tensor from a correspondingly oriented local into a
global coordinate system determines the exact form of the tensor
elements in the latter, in which the transport equation is usually
solved. This is of particular importance in the case of symmetry-free
magnetic fields, like the so-called Fisk field \citep{Fisk-1996}. The latter is --
although in a weaker manner than originally suspected
\citep{Lionello-etal-2006,Sternal-etal-2011} -- still a valid
generalisation of the Parker field and takes into account a
non-vanishing latitudinal field component.

While it has been recognized that the use of the Fisk field in models
of the heliospheric modulation of CRs requires a re-derivation of the
diffusion tensor \citep{Kobylinski-2001, Alania-2002,
  Burger-etal-2008} the formulas given in these papers differ from
each other and are either valid only for the case of isotropic
perpendicular diffusion (the former two papers) or for a specific
orientation of the local coordinate system (the latter paper).
Consequently, there are two open issues, namely (i) to determine which
of these formulas are correct (see also Appendix A) and (ii) to
generalize these results to the case of anisotropic perpendicular
diffusion.  With the present paper we address both issues by deriving
general formulas for the transformation of a fully anisotropic
diffusion tensor. In addition to establishing the appropriate
description, we apply the new generalized formulas to a standard
modulation problem in order to demonstrate the physical significance
of the approach.

\section{General considerations}
\begin{figure}[t!]
\centering
\includegraphics[width=0.95\columnwidth]{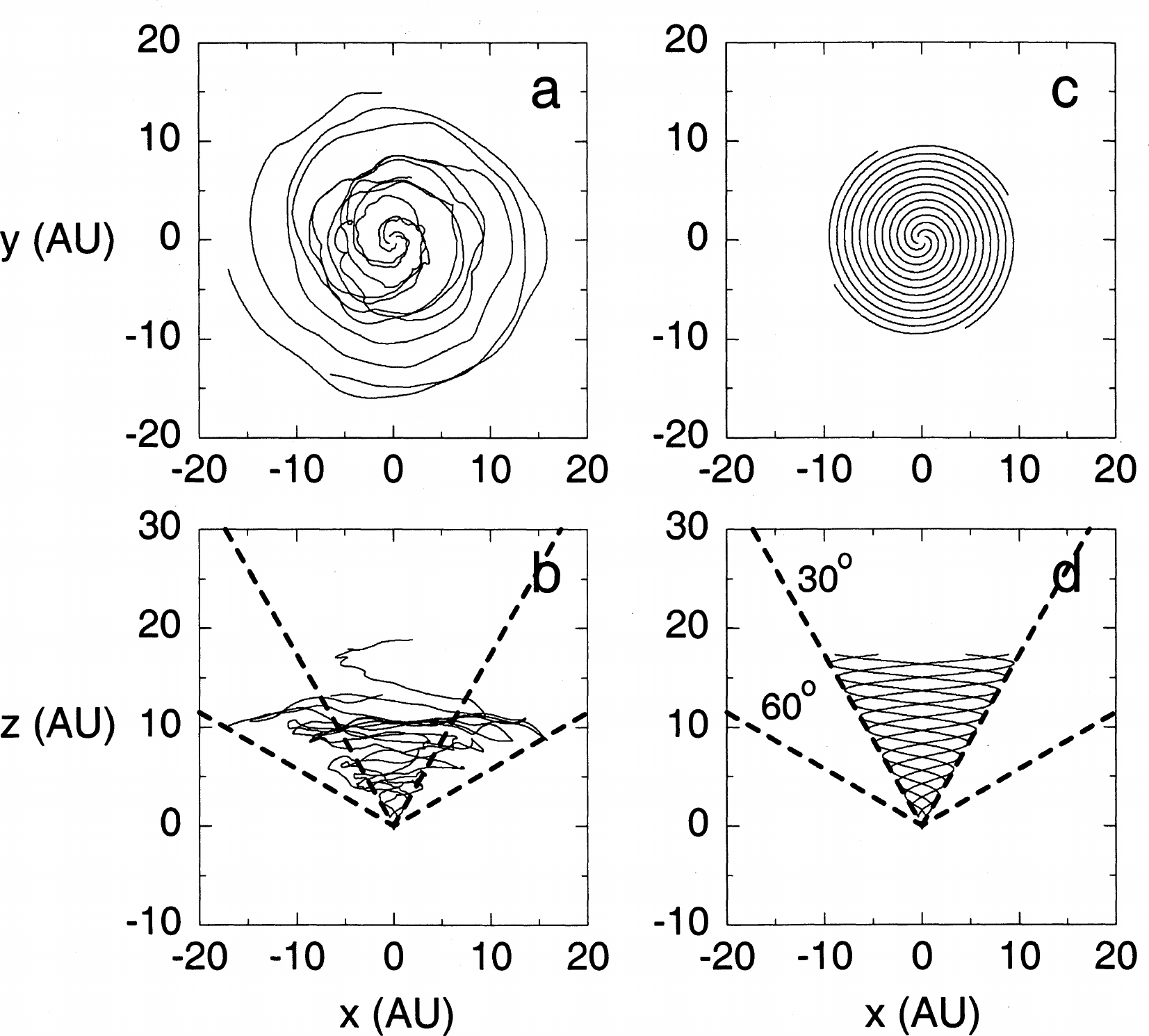}
\caption{The undisturbed (right) heliospheric magnetic field (projected into the 
         equatorial (top) and a meridional plane (bottom)) according to 
         \citet{Parker-1958} and its structure when field line random walk 
         is included (left), taken from \citet{Jokipii-2001}. 
         \label{fig:turb-hmf-parker}}
\end{figure}
Anisotropic perpendicular transport can, in principle, result (i) from
an inhomogeneous (asymmetric) magnetic background field or (ii) from
turbulence that is intrinsically non-axisymmetric with respect to the
(homogeneous) local magnetic field direction
\citep[e.g.,][]{Weinhorst-etal-2008}. While the latter case has been
discussed in the context of energetic particle transport
\citep{Ruffolo-etal-2008} partly motivated by the observed ratios of
the power in the microscale magnetic field fluctuations parallel and
perpendicular to the background field: $\delta B_{\perp 1}^2:\delta
B_{\perp 2}^2:\delta B_{\parallel}^2 = 5:4:1$ \citep[where $\delta
\vec{B}_{\perp 1}$ is aligned to the latitudinal unit vector and the
normalized $\delta \vec{B}_{\perp 2}$ completes the local trihedron,
see][]{Belcher-Davis-1971, Horbury-etal-1995}, recent analyses
indicate that the perpendicular fluctuations are probably axisymmetric
\citep{Turner-etal-2011, Wicks-etal-2012}. Therefore, we consider the
first case of an inhomogeneous magnetic background field to be more
likely to cause fully anisotropic diffusion.

If the random walk of field lines due to turbulence is significantly
contributing to the perpendicular particle transport, one generally
has to expect the latter to be anisotropic. This can be illustrated
already for the simple case that the HMF is represented by the Parker
spiral, see Fig.~\ref{fig:turb-hmf-parker}: Due to the field geometry
the field line wandering is not isotropic, neither in radial direction
nor in heliographic latitude, resulting in a field line diffusion
coefficient depending on both \citep{Webb-etal-2009}.

As soon as anisotropic perpendicular diffusion occurs, it is necessary
to determine the principal axes of the diffusion tensor in a local
field-aligned frame ($\kten_L$), because their orientation determines
the tensor elements in the global frame ($\kten$) after a
corresponding transformation given by
\begin{eqnarray}
\kten = A \kten_L A^{T} 
\end{eqnarray}
with 
\begin{align}
\kten_L=
\begin{pmatrix}
\kappa_{\perp 1} & \kappa_{A} & 0 \\
-\kappa_{A} & \kappa_{\perp 2} & 0 \\
0 & 0 & \kappa_{\parallel}
\end{pmatrix}
\end{align}
where, in general, $\kappa_A$ denotes the drift coefficient, induced
by a non-axisymmetric turbulence and by inhomogeneous magnetic
fields. The latter drifts can always be described by a drift velocity
$\vec{v}_d$ in the transport equation
\citep{Tautz-Shalchi-2012,Burger-etal-2008} and are therefore not
considered in the following. In Eq.~(1), analogous to the Euler angle
transformation known from classical mechanics, the matrix $A = R_3 R_2
R_1$ describes three consecutive rotations $R_i$ with $A^{-1} = A^T$.
These rotations are defined by the relative orientation of the local
and the global coordinate system with respect to each other.

Due to the latitudinal structuring of the solar wind and, in turn, of
the Parker spiral having a vanishing $B_{\vartheta}$-component, one
may argue that in that case the latitudinal direction remains a
preferred one so that the local coordinate system could always be
defined by the unit vectors $\vec{t}$ (along the field),
$\vec{e}_{\vartheta}$ (from a spherical polar coordinate system) and
$\vec{e}_{\vartheta} \times \vec{t}$. This, however, can obviously not
be the case for symmetry-free fields like the Fisk field \citep{Fisk-1996}.

In general, the local trihedron will consist of a unit vector
$\vec{t}$ tangential to the magnetic field and two orthogonal ones,
$\vec{u}$ and $\vec{v}$, defining the remaining principal axes. With
this notation the transformation (1) reads, for an arbitrary choice of
this local trihedron:
\begin{eqnarray}
\kappa_{11} & = & \kappa_{\perp 1}\,u_1^2    + \kappa_{\perp 2}\,v_1^2    + \kappa_{\parallel}\,t_1^2 \\
\kappa_{12} & = & \kappa_{\perp 1}\,u_1\,u_2 + \kappa_{\perp 2}\,v_1\,v_2 + \kappa_{\parallel}\,t_1\,t_2\\ & + & \kappa_A\,(u_1\,v_2-u_2\,v_1)\nonumber\\
\kappa_{13} & = & \kappa_{\perp 1}\,u_1\,u_3 + \kappa_{\perp 2}\,v_1\,v_3 + \kappa_{\parallel}\,t_1\,t_3\\ & + & \kappa_A\,(u_1\,v_3-u_3\,v_1)\nonumber\\
\kappa_{21} & = & \kappa_{\perp 1}\,u_1\,u_2 + \kappa_{\perp 2}\,v_1\,v_2 + \kappa_{\parallel}\,t_1\,t_2\\ & - & \kappa_A\,(u_1\,v_2-u_2\,v_1)\nonumber\\
\kappa_{22} & = & \kappa_{\perp 1}\,u_2^2    + \kappa_{\perp 2}\,v_2^2    + \kappa_{\parallel}\,t_2^2 \\
\kappa_{23} & = & \kappa_{\perp 1}\,u_2\,u_3 + \kappa_{\perp 2}\,v_2\,v_3 + \kappa_{\parallel}\,t_2\,t_3 \\ & + & \kappa_A\,(u_2\,v_3-u_3\,v_2)\nonumber\\
\kappa_{31} & = & \kappa_{\perp 1}\,u_1\,u_3 + \kappa_{\perp 2}\,v_1\,v_3 + \kappa_{\parallel}\,t_1\,t_3\\ & - & \kappa_A\,(u_1\,v_3-u_3\,v_1)\nonumber\\
\kappa_{32} & = & \kappa_{\perp 1}\,u_2\,u_3 + \kappa_{\perp 2}\,v_2\,v_3 + \kappa_{\parallel}\,t_2\,t_3 \\ & - & \kappa_A\,(u_2\,v_3-u_3\,v_2)\nonumber\\
\kappa_{33} & = & \kappa_{\perp 1}\,u_3^2    + \kappa_{\perp 2}\,v_3^2    + \kappa_{\parallel}\,t_3^2 
\end{eqnarray}
where the components of $\vec{t}, \vec{u}$ and $\vec{v}$ are determined in the global coordinate system.
Consequently, the task is to determine the unit vectors $\vec{t}$, $\vec{u}$, and $\vec{v}$ 
for an arbitrary, symmetry-free magnetic field.

Given that the perpendicular fluctuations are probably axisymmetric
\citep{Turner-etal-2011, Wicks-etal-2012} as discussed above, we
assume $\kappa_{A}=0$ in the following.

With this explicit formulation of the tensor elements we can already
address issue (i) defined in Section~1: For the case that the
perpendicular diffusion is isotropic, i.e.\ $\kappa_{\perp 1} =
\kappa_{\perp 2}$, the formulas given by \citet{Burger-etal-2008}, see
Appendix A, are identical to Eq.~(3) to (11), so that their correction
of the results found by \citet{Kobylinski-2001} and
\citet{Alania-2002} and, in turn, their subsequent analysis are
validated. We emphasize, however, that neither of these formulations
(involving only two rotation angles) allow to define explicitly the
perpendicular diffusion axes, which are nessesary to treat anisotropic
diffusion in the most general form.

\section{The choice of the local coordinate system} 
%\section{The local coordinate system} 
In the absence of symmetries there remain two distinguished local directions that, at a
given location within an arbitrary magnetic field, are related to its curvature $k$ and torsion
$\tau$ and are called the normal and the binormal direction. They can be defined with the
corresponding
normal and binormal unit vectors, respectively. Together with the tangential unit vector, they
constitute a local orthogonal trihedron fulfilling the ($k$- and $\tau$-defining) Frenet-Serret
relations \citep[e.g.,][]{Marris-Passman-1969}:
\begin{eqnarray}
\label{frenet-serret}
(\vec{t}\cdot\nabla)\vec{t} &=& k \vec{n}             \\
(\vec{t}\cdot\nabla)\vec{n} &=&-k \vec{t}+\tau\vec{b} \\
(\vec{t}\cdot\nabla)\vec{b} &=&-\tau\vec{n} 
\end{eqnarray}
If no other diffusion axes are preferred by any process, the
Frenet-Serret System constituted by the above definition of $\vec{t}$,
$\vec{n}$, and $\vec{b}$ is the most natural choice, i.e.\
$\vec{u}=\vec{n}$ and $\vec{v}=\vec{b}$ in Eqs. (3) to (11).

The transformation of the local diffusion tensor into a global
coordinate system according to these equations thus requires knowledge of the
dependence of the Frenet-Serret vectors on a given (non-homogeneous)
magnetic field $\vec{B}$. Evidently, the required relations are
\begin{eqnarray}
\vec{t}&=&\vec{B}/\vert\vec{B}\vert     \\
\vec{n}&=&(\vec{t}\cdot\nabla)\vec{t}/k \\
\vec{b}&=&\vec{t}\times\vec{n}          
\end{eqnarray}
This trihedron can, of course, only be established for a spatially non-homogeneous field, but 
this (weak) condition is fulfilled in most cases of interest. If there would
exist a region where the field is homogeneous, the choice of the vectors $\vec{n}$ and $\vec{b}$
is arbitrary (signifying isotropic perpendicular diffusion) unless 
no other preferential directions unrelated to the field geometry can be specified. Other 
principal directions unrelated to the large-scale geometry of the field could, for example, arise
from non-axisymmetric turbulence. The above formulas (3) to (11) remain unaffected, however: One only
needs to specify the appropriate vectors $\vec{t}, \vec{n}$ and $\vec{b}$. 

In the following we illustrate the procedure with the example of the
well-studied HMF. We quantitatively compare the new tensor
with the 'traditional' one, which is only valid for isotropic
perpendicular diffusion. This comparison reveals that a study of
fully anisotropic turbulent diffusion within more complicated fields
-- like the much-discussed heliospheric Fisk field
\citep{Burger-etal-2008,Sternal-etal-2011,Fisk-1996,
  Burger-Hitge-2004} or complex galactic magnetic fields
\citep{Ruzmaikin-etal-1988, Beck-etal-1996} -- has to be performed with
even more caution than thought before.

\begin{figure}[ht]
\centering
\includegraphics[width=0.95\columnwidth,angle=0]{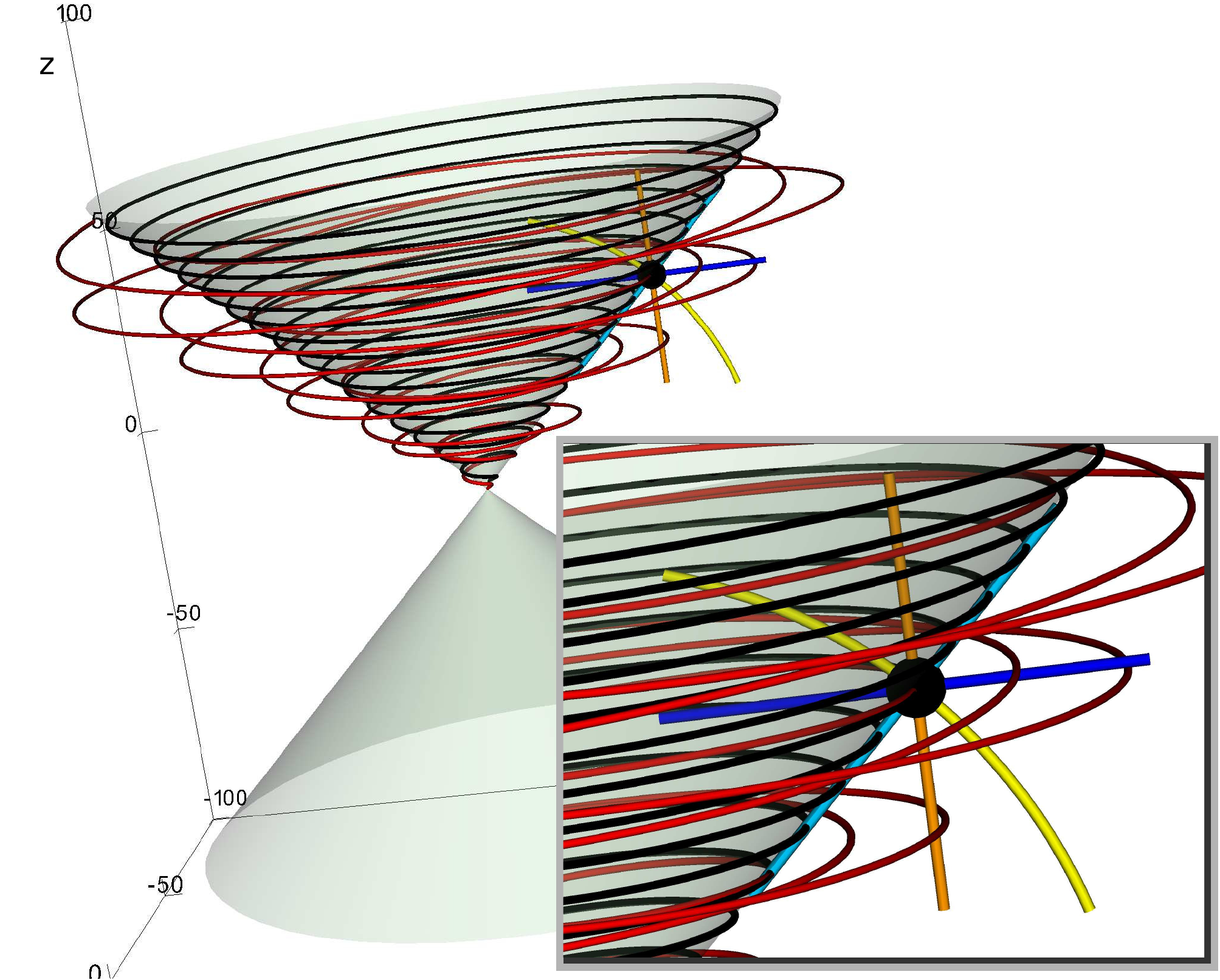}
\caption{The hybrid Fisk and the Parker field illustrated by red and
  black field lines, respectively. The two local trihedrons for the
  Parker field are indicated with the orange and blue (Frenet-Serret)
  as well as the yellow and light blue (traditional) lines.  Note that
  the traditional trihedron is always aligned to the Parker spiral
  cone of constant $\vartheta$ while for the Frenet-Serret trihedron
  one axis (the $\kappa_{\perp 2}$-binormal axis, orange) is
  nearly parallel to the $z$-direction. In the ecliptic both coordinate
  systems coincide by definition. All distances are in units of AU.\label{fig:trihed}}
\end{figure}

\section{An example for the new diffusion tensor}
\subsection{The heliospheric magnetic field}
An analytical representation of the HMF, which is referred to
as the {\it hybrid Fisk field}, can be found in \citet{Sternal-etal-2011}. 
For a constant solar wind speed ($u_{sw}=400$\,km/s) the HMF is represented, using spherical polar 
coordinates, by the following formulation:
\begin{align}
    \label{fisk}
    B_{r} &= A\,B_e\left(\frac{r_e}{r}\right)^2, \\
    B_{\vartheta} &= B_{r}\frac{r}{u_{sw}}\omega^*\sin\beta^* \sin\varphi^* \\
    B_{\varphi} &= B_{r}\,\frac{r}{u_{\rm{sw}}} 
    \left[\sin\vartheta(\omega^*\cos\beta^*-\Omega_\odot)     \right.\\\nonumber
    &+\left.\frac{d}{d\vartheta}(\omega^*\sin\beta^*\sin\vartheta)\cos\varphi^*\right.]
\end{align}
with
\begin{align}\nonumber
\beta^*(\vartheta)&=\beta F_{s}(\vartheta)\\\nonumber
\omega^*(\vartheta)&=\omega F_{s}(\vartheta)\\\nonumber
\varphi^* &= \varphi + \frac{\Omega_\odot}{u_{sw}}(r-r_{\odot})\nonumber
\end{align}
where
\begin{align}
&F_{s}(\vartheta) = \\
&\begin{cases}
[\tanh(\delta_p\vartheta) + \tanh(\delta_p(\vartheta-\pi))&\\ - \tanh(\delta_e(\vartheta-\vartheta_b))]^2 & 0 \le \vartheta < \vartheta_b \\
0 & \vartheta_b \le \vartheta \le \pi-\vartheta_b\\
[\tanh(\delta_p\vartheta) + \tanh(\delta_p(\vartheta-\pi))& \pi-\vartheta_b < \vartheta \le \pi\\ - \tanh(\delta_e(\vartheta-\pi+\vartheta_b))]^2\nonumber
\end{cases}
\end{align}
is the transition function introduced by \citet{Burger-etal-2008}. In
the case $F_s=0$ the HMF reduces to the standard Parker spiral
magnetic field.  For a quantitative comparison of different HMF
configurations, see \citet{Scherer-etal-2010}.

In Eq.~(18), $B_e$ denotes the magnetic field strength at $r_e=1$~AU,
$r_{\odot}$ is the solar radius, and $\Omega_{\odot} =
2.9\cdot10^{-6}$\,Hz is the averaged solar rotation frequency. The
constant $A= \pm 1$ in Eq.~(\ref{fisk}) indicates the different field
directions in the northern and southern hemisphere. The values for the
angle between the rotational and the so-called virtual axes of the Sun
$\beta = 12^{\circ}$ and the differential rotation rate $\omega =
\Omega_{\odot}/4$ are taken from \cite{Sternal-etal-2011}. The
parameters $\delta_{\rm{p}}=5$ and $\delta_{\rm{e}}= 5$ determine the
respective contributions of the Fisk- and Parker field above the poles
and in the ecliptic while $\vartheta_b=80^\circ$ is the cutoff
colatitude for the Fisk-field influence. In the following, we consider
two cases, a pure Parker field (i.e.\ setting $F_{s}=0$ in
Eqs.~(19) and (20)) and the hybrid Fisk field with $F_s$ from Eq.~(21).
Both fields are illustrated by exemplary field lines in Fig.~\ref{fig:trihed}.

\subsection{The local diffusion tensor elements}
The elements of the local diffusion tensor are chosen following the approach in
\citet{Reinecke-etal-1993}, i.e.\ as
\begin{align}
\kappa_{\parallel} &=\kappa_{\parallel_0}\beta\left(\frac{p}{p_0}\right)
\left(\frac{B_e}{B}\right)^{a_\parallel}\label{eq:kappapar}\\
\kappa_{\perp 1} &=\kappa_{\perp_0}\beta\left(\frac{p}{p_0}\right)
\left(\frac{B_e}{B}\right)^{a_\perp}\label{eq:kappaperp1}\\
\kappa_{\perp 2} &=\xi\kappa_{\perp 1}\label{eq:kappaperp2}
\end{align}
where $\beta=v/c$ is the particle speed normalized to the speed of light,
$p$ is the particle momentum with the normalisation constant
$p_0=1$\,GeV/c and $B$ is the magnitude of the magnetic field.  The
scaling exponents have the values $a_\parallel = 0.75$ and
$a_\perp=0.97$. The parallel diffusion constant is
$\kappa_{\parallel_0}=0.9\cdot 10^{22}$cm$^2$/s while
$\kappa_{\perp_0}=0.1\kappa_{\parallel_0}$. The anisotropy in
perpendicular diffusion is assumed to be soley determined by the
factor $\xi$ which is set to 2 for the following discussion. This is
still a moderate choice compared to the findings of, e.g.,
\citet{Potgieter-etal-1997}.

\begin{figure}[ht]
\centering
\includegraphics[width=0.95\columnwidth,angle=0]{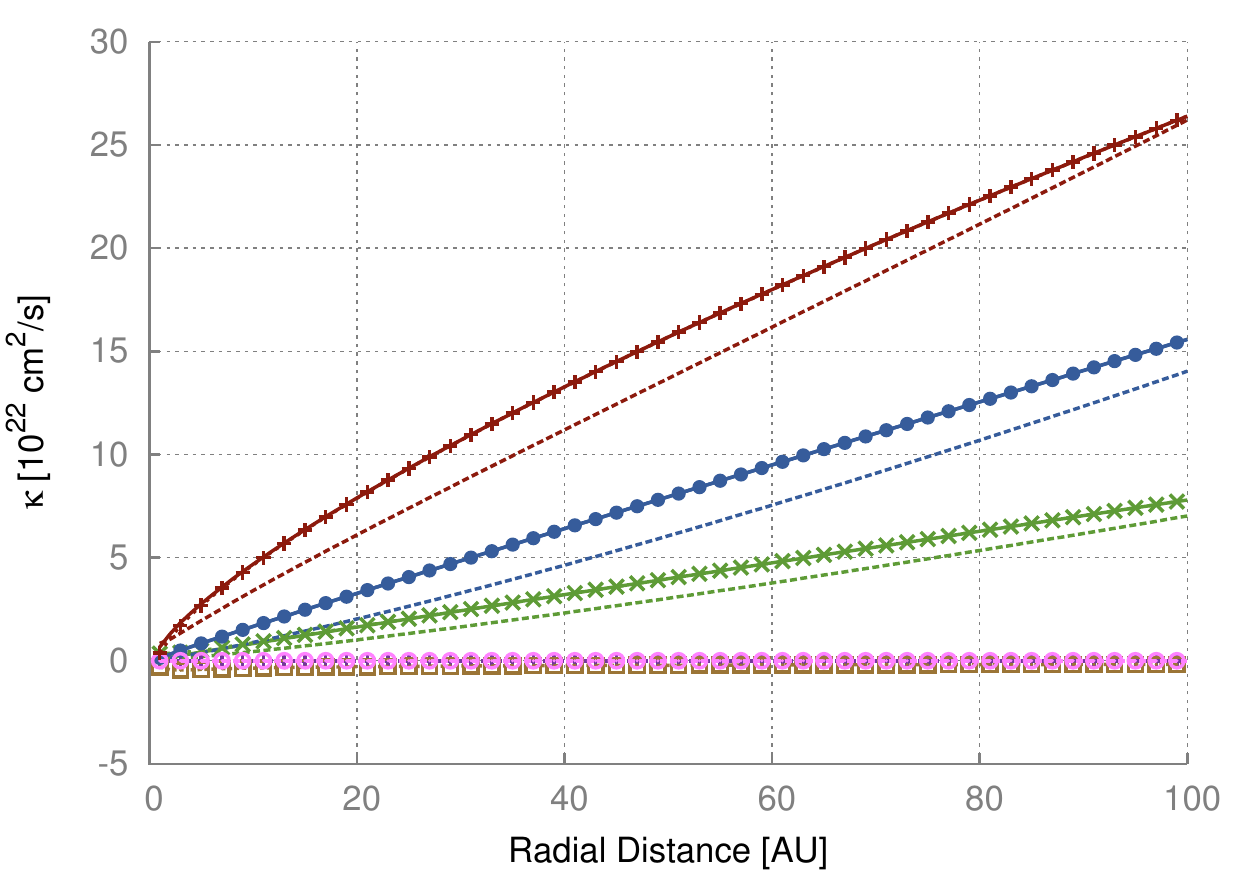}
\caption{The dependence of the local and global tensor elements on
  heliocentric distance in the ecliptic plane for the Parker
  field. The local elements from the formulas
  (\ref{eq:kappapar})-(\ref{eq:kappaperp2}) are shown as solid red
  ($\kappa_{\parallel}$), green ($\kappa_{\perp 1}$), and blue
  ($\kappa_{\perp 2}$) curves. The results for the parallel diffusion
  coefficent $\kappa_\parallel^{(\rm{QLT})} =
  1/3v\lambda_{\parallel}^{(\rm{QLT})}$ (Eq.\ref{eq:QLT}) and the
  perpendicular coefficient from $\kappa_\perp^{(\rm{NLGC})}$
  (Eq.~\ref{eq:NLGC}) are drawn as dashed red and green lines,
  respectively, while the dashed blue curve is just scaled as
  $\xi\kappa_\perp^{(\rm{NLGC})}$ with $\xi=2$. The overlayed,
  color-matched symbolds show the nearly perfect alignment of the
  $\kappa_{rr}$ (green, $\times$), $\kappa_{\vartheta\vartheta}$ (blue,
  $\bullet$) and $\kappa_{\varphi\varphi}$ (red, $+$) global diagonal
  tensor elements in the ecliptic, due to the Parker field
  structure. All other tensor elements are almost indistinguishable
  from zero, as indicated by the remaining
  symbols.\label{fig:radtens}}
\end{figure}

Although these empirical formulas for the local diffusion coefficients
are not directly related to the turbulence evolution in the
heliosphere and more sophisticated theoretical models for the
corresponding mean free paths in parallel and perpendicular direction
exist, they are still a good approximation as can be seen in the
following.  The result from quasilinear theory (QLT) for the mean
free path (see e.g. \citet{Shalchi-2009}) is given by
\begin{align}
\lambda&_{\parallel}^{(\rm{QLT})} =\nonumber\\ &\frac{3 l_{\rm{slab}}}{16\pi C(\nu)}\left(\frac{B}{\delta B_{\rm{slab}}}\right)^2 R^{2-2\nu}\left[\frac{2}{(1-\nu)(2-\nu)} + R^{2\nu}\right]
\label{eq:QLT}
\end{align}
with
\begin{equation}
C(\nu) = \frac{1}{2\sqrt{\pi}}\frac{\Gamma(\nu)}{\Gamma(\nu-1/2)}
\end{equation}
where $\Gamma(x)$ is the gamma function, $2\nu=5/3$ is the inertial
range spectral index, $R=R_L/l_{slab}$ is the dimensionless rigidity,
and $R_L = pc/(|q|B)$ is the particle Larmor radius. If one
scales the bendover scale of slab turbulence as
$l_{slab}=0.03\,\rho^{0.5}$ (where $\rho$ is the heliocentric distance
in AU) and the slab turbulence variance as $\delta B_{slab}^2 =
B_{e}^2 \rho^{-2.15}$, the radial dependence of the local tensor
elements matches well with the approximative formulas
(\ref{eq:kappapar}) to (\ref{eq:kappaperp2}) as shown in 
Fig.~\ref{fig:radtens}. It is interesting to note that these scalings are
similar to the assumptions made in \citet{Burger-etal-2008}. They use
the same radial dependence for $l_{slab}$ (their exponent of
$1/l_{slab}=k_{min}=32\,\rho^{0.5}$ is a typing error, private
communication with the authors) and an exponent of $-2.5$ for the slab
turbulence variance $\delta B_{slab}^2$, which is slightly larger.
Similar arguments can be made for the perpendicular
diffusion. Employing for the perpendicular diffusion coefficient the
result of the nonlinear guiding center (NLGC) theory of
\citet{Matthaeus-etal-2003} and \citet{Shalchi-etal-2004}, namely
\begin{align}
\kappa&_{\perp}^{(\rm{NLGC})} =\nonumber\\ &\left[a^2v\frac{\nu-1}{2\sqrt{3}\nu}\sqrt{\pi}\frac{\Gamma(\nu/2 + 1)}{\Gamma(\nu/2 + 1/2)}l_{2D}\frac{\delta B_{2D}^2}{B^2}\right]^{2/3}\kappa_{\parallel}^{1/3}
\label{eq:NLGC}
\end{align}
(see formula (15) in \citet{Burger-etal-2008}) with the constant
$a=1/\sqrt{3}$. Scaling again the 2D turbulence correlation length
$l_{2D}$ with $\rho^{0.5}$ and the turbulence variance $\delta
B_{2D}^2$ even more weakly with $\rho^{-1.2}$ yields results similar
to those obtained by \citet{Reinecke-etal-1993} for the perpendicular
diffusion, as shown in Fig.~\ref{fig:radtens} as well.

Given the uncertainties both in the actual magnetic turbulence
evolution in the heliosphere with radial distance and latitude (see,
e.g., \citet{Oughton-etal-2011} for a study in which they find a much
more complicated radial dependence of the slab variance) and their
actual relation to perpendicular or even anisotropic perpendicular
diffusion in connection with three-dimensional turbulence
\citep{Shalchi-2010,Shalchi-etal-2010b} we stick, in the following,
with the empirical formulas of Eqs.~(\ref{eq:kappapar}) to
(\ref{eq:kappaperp2}) for this principal study.

\subsection{The structure of the global diffusion tensor}
Employing the formalism described in Section 3 to calculate the global
diffusion tensor \kten\ results in tensor elements $\kappa_{ij}$ which
are different from those 'traditionally' used, labeled
${\kappa}^B_{ij}$ here, with $i,j\in\{r,\vartheta,\varphi\}$. The
latter are derived following the transformation presented by
\citet{Burger-etal-2008}, which for the Parker field is equivalent to
the assumption that the local system can always be defined by
$\vec{t}$, $\vec{n}=\vec{e}_{\vartheta}\times\vec{t}$, and
$\vec{b}=\vec{e}_{\vartheta}$, see Appendix A for the detailed
transformation formulas.  Both local systems are illustrated in
Fig.~\ref{fig:trihed}.

\begin{figure*}[t]
\centering
\includegraphics[width=0.98\textwidth]{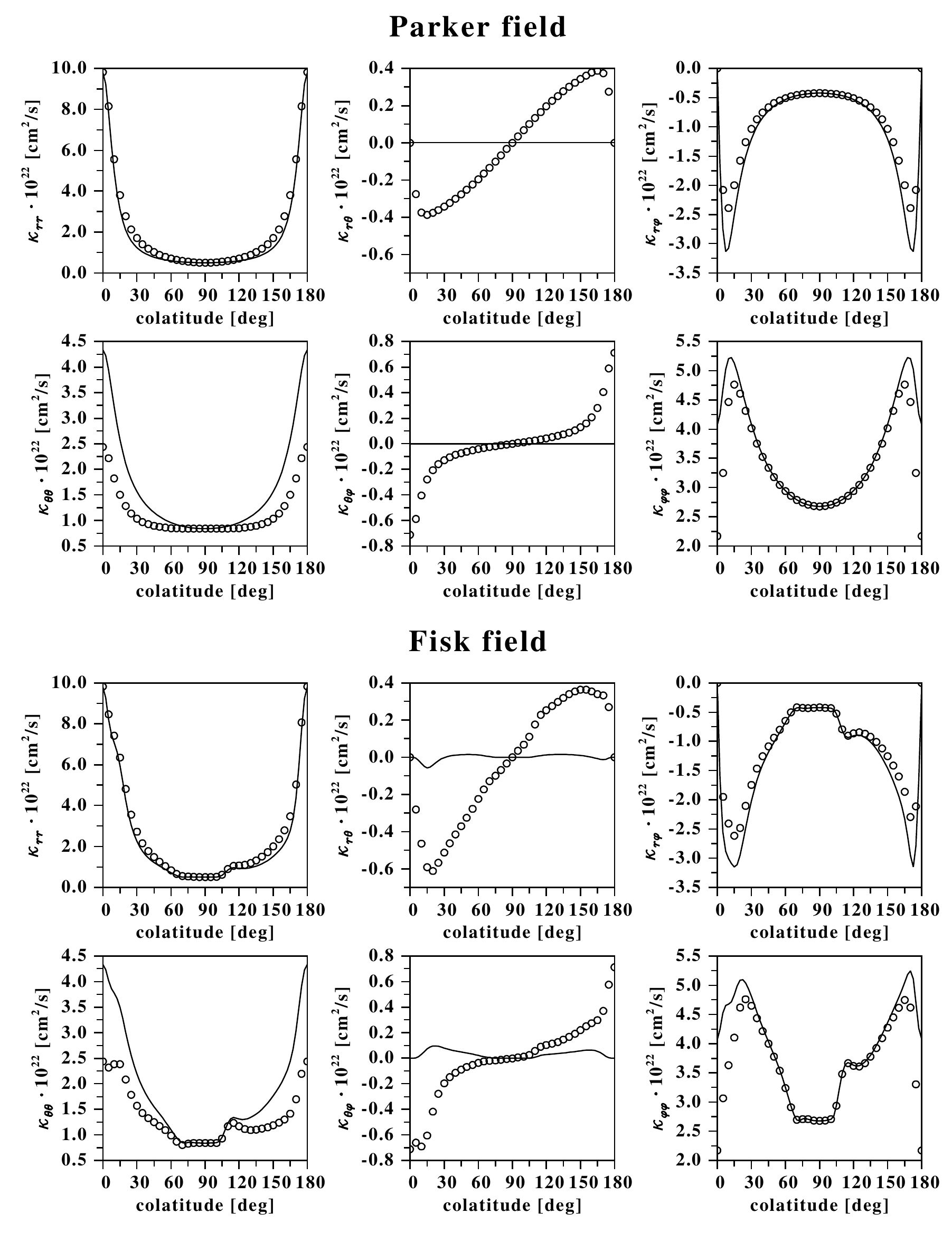}
\caption{The six independent tensor elements $\kappa_{ij}^B$ for the
  ``traditional'' tensor formulation following
  \citet{Burger-etal-2008} (solid line) and the new $\kappa_{ij}$
  using the Frenet-Serret trihedron (open circles) for a fixed radius
  of $r=5$ AU, a longitude of $\varphi=\pi/4$ and for varying
  colatitude. The upper two rows show those for the Parker field while
  the lower two rows show those for the hybrid Fisk
  field.\label{fig:panel}}
\end{figure*}

The different behavior of the tensor elements $\kappa_{ij}$ and
${\kappa}^B_{ij}$ with latitude at a heliocentric distance of 5~AU and
longitude $\varphi=\pi/4$ is displayed in Fig.~\ref{fig:panel} for the
Parker and hybrid Fisk field, respectively. By definition, both
formulations yield the same tensor elements in the ecliptic plane,
i.e. for $\vartheta=\pi/2$, while for higher latitudes the
differences become more and more pronounced.

In the Parker case, e.g.\ the upper two rows in Fig.~\ref{fig:panel},
the elements $\kappa_{rr}$, $\kappa_{r\varphi}$, and
$\kappa_{\varphi\varphi}$ show roughly the same behavior for all
latitudes. The strongest mixing of the local elements $\kappa_{\perp 1}$
and $\kappa_{\perp 2}$ occurs in $\kappa_{\vartheta\vartheta}$,
so that the deviations for high latitudes are more
pronounced. The main difference appears in the off-diagonal elements
$\kappa_{r\vartheta}$ and $\kappa_{\vartheta\varphi}$, which are
different from zero in the general case discussed here while they are
equal to zero in the traditional approach.

The differences between the hybrid Fisk field tensor elements (shown
in the lower two rows in Fig.~\ref{fig:panel}) are similar to those of
the Parker field described above, although they show a more
complicated $\vartheta$ dependence. Note that in the traditional
formulation the off-diagonal elements $\kappa_{r\vartheta}$ and
$\kappa_{\vartheta\varphi}$ are already nonzero for the hybrid Fisk
field and become larger in the new formulation.

\begin{figure}[t]
\centering
\includegraphics[width=0.95\columnwidth]{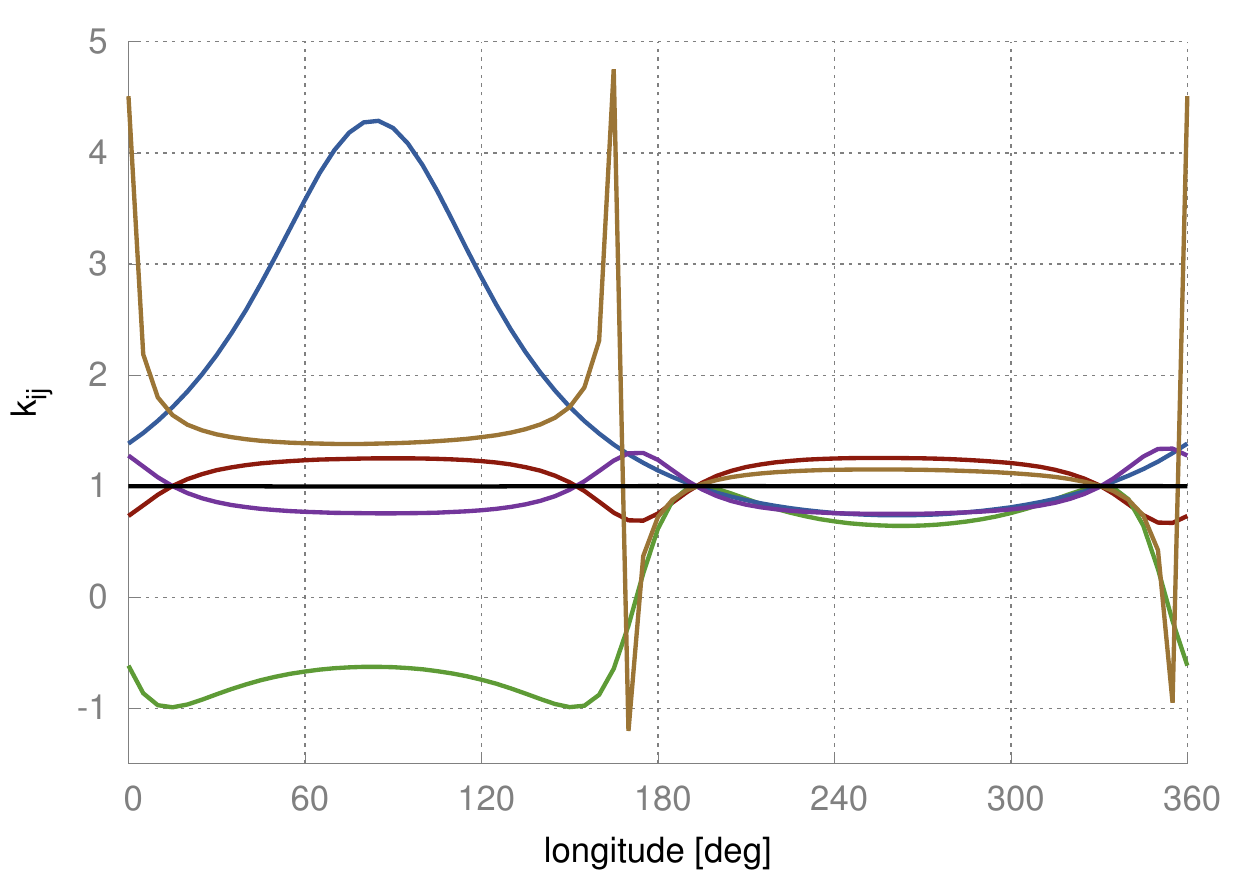}
\caption{Ratios of the tensor elements
  $k_{ij}=\kappa_{ij}^B/\kappa_{ij}$ for the hybrid Fisk field plotted
  against heliographic longitude for a fixed heliocentric distance of
  $r=50$ AU and a heliographic colatitude of $\vartheta = \pi/4$. The
  elements shown are $k_{rr}$ (red), $k_{r\vartheta}$ (green),
  $k_{r\varphi}$ (blue), $k_{\vartheta\vartheta}$ (violett),
  $k_{\vartheta\varphi}$ (brown), and $k_{\varphi\varphi}$
  (black).\label{fig:longratios}}
\end{figure}
%
%\clearpage

The choice of longitude is arbitrary for the Parker field, since it
has no $\varphi$ dependence.  The Fisk field, however, has significant
longitudinal variations, therefore, we show the ratios of the
traditional and the new tensor elements for the hybrid Fisk field with
longitude (Fig.~\ref{fig:longratios}). It can be seen that for large
heliocentric distances, the deviations between both formulations vary
strongly, illustrated here for a heliocentric distance of $r=50$ AU and
a heliographic colatitude of $\vartheta=\pi/4$.

We emphasize again that in the case of isotropic perpendicular diffusion ($\xi=1$),
the traditional and the new formulations are identical for any given magnetic 
field with non-vanishing curvature. The differences between them scale with the
perpendicular anisotropy $\xi$ (see Eq. \ref{eq:kappaperp2}). 

% %
% \begin{figure}[ht]
% \epsscale{1.}
% \plotone{fig2}
% \caption{The local trihedron for the Fisk field, illustrated by field
%   lines. The field line of the Fisk magnetic field is plotted in red
%   while the Parker field is in black. Yellow and violet represent the
%   calculated normal and binormal directions, respectively.  The thin
%   blue and orange lines indicate the $\vec{e}_\vartheta$ and
%   $\vec{e}_\vartheta\times\vec{t}$ from the 'traditional' trihedron
%   (all distances in units of AU, {1 AU}$=1.49\cdot
%   10^{11}$m).\label{fig:trihed}}
% \end{figure}
% %
\section{Application to the modulation of cosmic ray spectra}
\begin{figure}[t]
\centering
%\epsscale{1.0}
%\plottwo{fig5}{fig6}
\includegraphics[width=0.21\textwidth]{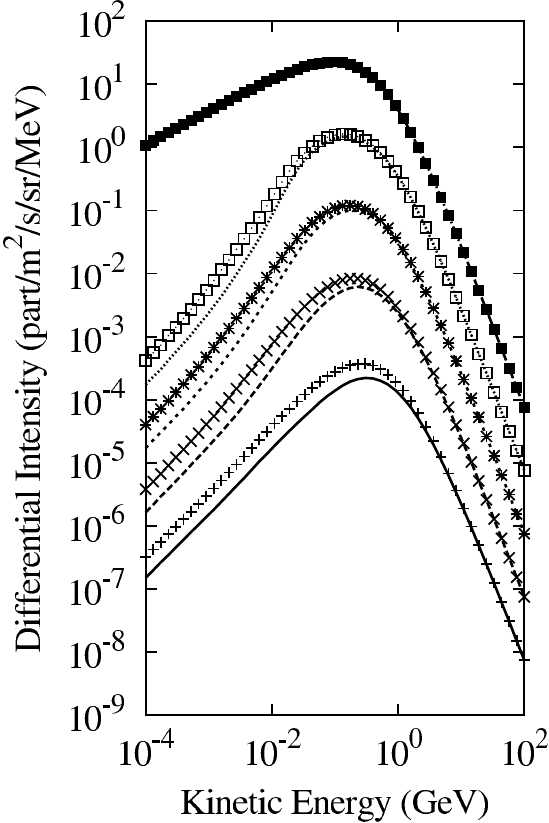}
\includegraphics[width=0.219\textwidth]{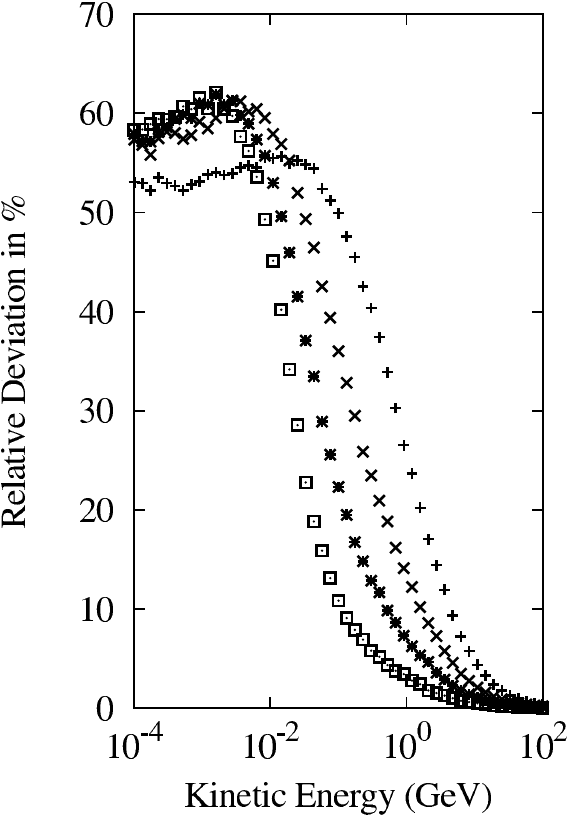}
\caption{Modulated spectra for fully anisotropic diffusion of galactic
  protons for both tensor formulations. The left panel shows the
  resulting spectra for four heliospheric distances (1~AU, 25~AU,
  50~AU, 75~AU, from bottom to top) and the LIS modulation boundary at
  100~AU (solid squares). The spectra, shifted in the plot by powers
  of 10 for clarity (note the resulting high energy offsets), converge
  to the LIS for high energies. While the symbols indicate the results
  from the new tensor formulation with the Frenet-Serret orientation,
  the lines are results from an analogous computation employing the
  'traditional' two-angle transformation. In both cases is
  $\kappa_{\perp 2}=2\,\kappa_{\perp 1}$. The right panel gives the
  relative deviations (normalized to the new results) of corresponding
  spectra from each other. The symbols are the same as in the left
  panel.\label{fig:spectra}}
\end{figure}
To assess the impact of the new tensor formulation on CR modulation we
employ a CR proton transport model by solving the Parker equation
\citep{parker-1965}
\begin{equation}
\label{eq:cr}
\frac{\partial f}{\partial t} =
\nabla\cdot(\kten\nabla f) -
\vec{u}_s\cdot\nabla f + \frac{p}{3}\,(\nabla\cdot \vec{u}_s)\,\frac{\partial f}{\partial p}
\end{equation}
to determine the differential CR intensity $j(\vec{r},p,t) = p^2
f(\vec{r},p,t)$ (with $\vec{r}$ as the position in three-dimensional
configuration space and $p$ as momentum). The solar wind velocity $\vec{u}_s$ is
radially pointing outwards with a constant speed of 400\,km/s and
\kten\ is the diffusion tensor in the global frame for the Parker
spiral magnetic field. This implies the Frenet-Serret trihedron of the
form explicitly given in Appendix B.

The solution is obtained via a numerical integration of an equivalent
system of stochastic differential equations (SDEs)
\citep{Kopp-etal-2011,Gardiner-1994}
\begin{equation}
dx_i = A_i(x_i)\,dt + \sum_j B_{ij}(x_i)\,dW_j
\end{equation}
for an ensemble of pseudo-particles (phase space elements) with
$\kten=\hat{B}{\hat{B}}^T$ and $d{\vec{W}}(t)=\sqrt{dt}\,\vec{N}(t)$ where
${\vec N}(t)$ is a vector of normal distributed random numbers and
$x_i$ denotes the phase space coordinates. The stochastic motion
$d{\vec{W}}(t)$ is often refereed to as Wiener process. The deterministic
processes from Eq.~(\ref{eq:cr}) like the advection with the solar
wind flow and the adiabatic energy changes are contained in the
generalized velocity $\vec{A}$. We employ the time-backward Markov
stochastic method, meaning that we trace back the pseudo-particles
from a given phase space point of interest, until they hit the
integration boundary. The solution to the transport equation
(\ref{eq:cr}) is then constructed as a proper average over the
pseudo-particle orbits. For details on the general method and the
numerical scheme, especially in the case of a general diffusion
tensor, see \citet{Kopp-etal-2011}, \citet{Strauss-etal-2011a}, and
\citet{Strauss-etal-2011b}.

The local interstellar spectrum (LIS) of protons $j_{LIS}$ is assumed
at a heliocentric distance of $r=100$~AU as a spherically symmetric
Dirichlet boundary condition. At the inner boundary of one solar
radius $r=R_\sun$ the pseudo-particles are reflected, which is
equivalent to a vanishing gradient in the CR density there. A
standard representation of the proton LIS is given by
\begin{equation}
j_{LIS} = 12.14\,\beta(E_{kin} + 0.5E_{0})^{-2.6}
\end{equation}
and was taken from \citet{Reinecke-etal-1993}. The proton rest energy
$E_{0}$ is equal to 0.938 and $E_{kin}$ denotes the kinetic energy of
a particle (both in units of GeV).

The LIS and the resulting modulated spectra are shown in
Fig.\ref{fig:spectra} for both tensor formulations and for several
heliocentric distances. The spectra for the new Frenet-Serret tensor are
higher by up to 60\% at low energies  for all heliocentric
distances. This is due to the enhanced diffusive flux from the
modulation boundary via an effective inward diffusion along the
polar axis. In the tensor formulation provided by \citet{Burger-etal-2008} this
diffusion (determined by $\kappa_{\perp 2}$) cannot transport particles from the
boundary into the inner heliosphere, it merely distributes the particles
on a shell of fixed heliocentric distance. In the new tensor formulation
exists thus a form of 'pseudo drift' produced by the off-diagonal
tensor elements in the global frame, which were different or even equal to zero in
the traditional formulation. This reduced modulation effect is
relevant for higher energies at lower heliocentric distances,
since the particles have more time to adiabatically cool (see the right panel
of Fig. \ref{fig:spectra}).

\newpage
\section{Conclusions}
We have derived, in a global reference frame, the general form of the
diffusion tensor of energetic particles in arbitrary magnetic
fields. This new formulation particularly includes the case of
anisotropic perpendicular diffusion that arises from field line
wandering or scattering due to turbulence and requires a determination
of both principal (orthogonal) perpendicular diffusion directions.
Unless the turbulence is non-axisymmetric, which appears to be
unlikely for the solar wind, the natural choice for these principal
directions is the Frenet-Serret trihedron associated with the
curvature and torsion of the magnetic field lines.

After the derivation of the formulas for all tensor elements in
dependence of the Frenet-Serret unit vectors, we have first
quantitatively compared the results to those published previously for
the example of the heliospheric magnetic field. For the latter we have
discussed two well-established alternatives, namely the Parker field
and the hybrid Fisk field. While the old and new tensor formulations
coincide for the case of isotropic perpendicular diffusion, the more
general case of anisotropic perpendicular diffusion cannot be treated
consistently with the earlier approaches. This is manifest in
significant differences of corresponding tensor elements including
additional non-vanishing ones.

Second, we have demonstrated the consequences of the new
tensor formulation in application to the modulation of galactic
cosmic ray proton spectra in the Parker heliospheric magnetic
field. Solving the cosmic ray transport equation with the method of
stochastic differential equations allowed us to quantify the
differences between the spectra resulting from both tensor
formulations for the case of perpendicular diffusion with an
anisotropy of $\xi=2$. We found those differences to amount up to 60\%
at energies below a few hundred MeV. Given that we used for this first
principal assessment an anisotropy that is moderate as compared to
findings from detailed transport and modulation studies, the fluxes
can be influenced even more strongly and at even higher energies.

Besides the fact that the above results indicate the necessity to
study the case of fully anisotropic diffusion in more detail within
the framework of more sophisticated models of heliospheric cosmic ray
modulation, they can furthermore be expected to be of importance for
the particle transport in complex galactic magnetic fields for which
usually isotropic (scalar) diffusion has been considered so far.

\acknowledgments{\small Acknowledgements: The work was carried out
  within the framework of the `Galactocauses' project (FI 706/9-1)
  funded by the Deutsche Forschungsgemeinschaft (DFG) and benefitted
  from the DFG-Forschergruppe FOR 1048 (project FI 706/8-1/2), the
  'Heliocauses' DFG-project (FI 706/6-3) as well as from the project
  SUA08/011 financed by the Bundesministerium f\"ur Forschung und
  Bildung (BMBF). We thank I.\ B\"usching and A.\ Kopp for providing
  the basis for the SDE numerical solver. We also thank N.E.~Engelbrecht
  for helpfull discussions and an anonymous referee for a constructive
  evaluation.}

\begin{appendix}
\section{Burger Transformation Formulas}
The transformation formulas for the diffusion tensor given in \citet{Burger-etal-2008} read:
\begin{align}
\kappa_{rr}^B &= \kappa_{\perp 2}\sin^2\zeta + \cos^2\zeta (\kappa_{\parallel}\cos^2\Psi + \kappa_{\perp 1}\sin^2\Psi) \nonumber\\
\kappa_{r\vartheta}^B &= \sin\zeta\cos\zeta(\kappa_{\parallel}\cos^2\Psi + \kappa_{\perp 1}\sin^2\Psi - \kappa_{\perp 2}) \nonumber\\
\kappa_{r\varphi}^B &= -\sin\Psi\cos\Psi\cos\zeta(\kappa_{\parallel} - \kappa_{\perp 1}) \nonumber\\
\kappa_{\vartheta\vartheta}^B &= \kappa_{\perp 2}\cos^2\zeta + \sin^2\zeta (\kappa_{\parallel}\cos^2\Psi + \kappa_{\perp 1}\sin^2\Psi) \nonumber\\
\kappa_{\vartheta\varphi}^B &= -\sin\Psi\cos\Psi\sin\zeta(\kappa_{\parallel} - \kappa_{\perp 1}) \nonumber\\
\kappa_{\varphi\varphi}^B &= \kappa_{\parallel}\sin^2\Psi + \kappa_{\perp 1}\cos^2\Psi
\end{align}
with $\tan\Psi = -B_\varphi/\sqrt{B_r^2 + B_\vartheta^2}$ and
$\tan\zeta = B_{\vartheta}/B_r$. Note that \citet{Kobylinski-2001} and
\citet{Alania-2002} state a different formula for $\Psi$, namely
$\tan\Psi = -B_\varphi/B_r$. Moreover, these formulas involve only two
angles in contrast to the general case described with the matrix $A$
in Eq.~(1) in Section~2. As discussed in the text,
these formulas in the given form can only hold for $\kappa_{\perp 1} \neq \kappa_{\perp
  2}$ in case of special magnetic fields with $B_{\vartheta}=0$ like
that introduced by Parker.

\section{The Parker Frenet-Serret Trihedron}
Here, we derive the analytic expressions for the Frenet-Serret trihedron for
the Parker case of the HMF. Reducing Eqs.~(18) to (20) to the Parker field 
by setting $F_s=0$ we obtain
\begin{equation}
\label{tangentialvec}
\vec{t} = \frac{\vec{e}_r - \tan \chi \vec{e}_\varphi}{\sqrt{1+\tan^2 \chi}}
\end{equation}
for the tangential vector, with $\displaystyle\tan \chi = \frac{\omega}{u_{s}} r
\sin\vartheta$. The easiest way to derive the normal vector $\vec{n}$
is to calculate $(\vec{t}\cdot\nabla)\vec{t} = k \vec{n}$ (where $k$
is the curvature, see Eq.(\ref{frenet-serret})) and to normalize
appropriately. After some straightforward calculation, one arrives at
\begin{equation}
\label{normalvec}
\vec{n} = -\frac{E(\tan\chi\vec{e}_r + \vec{e}_\varphi) + F \vec{e}_\vartheta}{\sqrt{E^2(\tan^2\chi + 1) + F^2}}
\end{equation}
where the abbrevations
\begin{equation}
\label{abbrev}
E= \frac{\tan^2\chi}{r} + \frac{\omega}{u_{s}}\frac{\sin\vartheta}{1+\tan^2\chi}, \qquad
F=\frac{\tan^2\chi\cos\vartheta}{r\sin\vartheta}
\end{equation}
have been introduced. The binormal vector is now simply the cross product $\vec{b} = \vec{t}\times\vec{n}$, which yields
\begin{equation}
\label{binormalvec}
\vec{b} = \frac{-F(\tan\chi\vec{e}_r + \vec{e}_\varphi) + E(\tan^2\chi + 1)\vec{e}_\vartheta}{\sqrt{F^2(\tan^2\chi + 1) + E^2(\tan^2\chi + 1)^2}}
\end{equation}
Eqs.(B1), (B2), and (B4) are the explicit formulas for the
Frenet-Serret trihedron in the case of the heliospheric Parker
field. Corres~ponding but much longer expressions can, in principle,
be obtained for the Fisk field as well.
\end{appendix}

%\bibliography{references}
%\clearpage
%\end{document}

\end{document}